\documentclass[aps,prl,reprint,superscriptaddress,showpacs,altaffilletter]{revtex4-1}
\usepackage{graphicx}
\usepackage{multirow}
\usepackage{color}
\usepackage{amsmath, amssymb, amsfonts, amsthm}
\usepackage{dcolumn}
\usepackage[table]{xcolor}
\usepackage[caption=false]{subfig}

\usepackage{hyperref}

\begin{document}

\title{Measurement of the half-life of the T=$\frac{1}{2}$ mirror decay of $^{19}$Ne and its implication on physics beyond the standard model}

\author{L.J. Broussard}
\altaffiliation[Present address: ]{Los Alamos National Laboratory, Los Alamos, NM 87544  USA}
\altaffiliation[]{leahb@lanl.gov}
\affiliation{Duke University, Durham, NC 27708  USA}
\affiliation{Triangle Universities Nuclear Laboratory (TUNL), Durham, NC, 27708  USA}
\author{H.O. Back}
\altaffiliation[Present address: ]{Princeton University, Princeton, NJ  08544  USA}
\affiliation{Triangle Universities Nuclear Laboratory (TUNL), Durham, NC, 27708  USA}
\affiliation{North Carolina State University, Raleigh, NC 27695  USA}
\author{M.S. Boswell}
\altaffiliation[Present address: ]{Los Alamos National Laboratory, Los Alamos, NM 87544  USA}
\affiliation{Triangle Universities Nuclear Laboratory (TUNL), Durham, NC, 27708  USA}
\affiliation{University of North Carolina, Chapel Hill, NC 27599  USA}
\author{A.S. Crowell}
\affiliation{Duke University, Durham, NC 27708  USA}
\affiliation{Triangle Universities Nuclear Laboratory (TUNL), Durham, NC, 27708  USA}
\author{P. Dendooven}
\affiliation{University of Groningen, KVI, 9747 AA Groningen, The Netherlands}
\author{G.S. Giri}
\altaffiliation[Present address: ]{Department of Physics and Astronomy, University of Sussex, Brighton BN1 9QH, United Kingdom}
\affiliation{University of Groningen, KVI, 9747 AA Groningen, The Netherlands}
\author{C.R. Howell}
\affiliation{Duke University, Durham, NC 27708  USA}
\affiliation{Triangle Universities Nuclear Laboratory (TUNL), Durham, NC, 27708  USA}
\author{M.F. Kidd}
\altaffiliation[Present address: ]{Tennessee Tech University, Cookeville, TN 38505  USA}
\affiliation{Duke University, Durham, NC 27708  USA}
\affiliation{Triangle Universities Nuclear Laboratory (TUNL), Durham, NC, 27708  USA}
\author{K. Jungmann}
\altaffiliation[Present address: ]{University of Groningen, FMNS, 9747 AG Groningen, The Netherlands}
\affiliation{University of Groningen, KVI, 9747 AA Groningen, The Netherlands}
\author{W.L. Kruithof}
\affiliation{University of Groningen, KVI, 9747 AA Groningen, The Netherlands}
\author{A. Mol}
\affiliation{University of Groningen, KVI, 9747 AA Groningen, The Netherlands}
\author{C.J.G. Onderwater}
\altaffiliation[Present address: ]{University of Groningen, FMNS, 9747 AG Groningen, The Netherlands}
\affiliation{University of Groningen, KVI, 9747 AA Groningen, The Netherlands}
\author{R.W. Pattie, Jr.}
\affiliation{Triangle Universities Nuclear Laboratory (TUNL), Durham, NC, 27708  USA}
\affiliation{North Carolina State University, Raleigh, NC 27695  USA}
\author{P.D. Shidling}
\altaffiliation[Present address: ]{Cyclotron Institute, Texas A$\&$M University, College Station, TX, 77843  USA}
\affiliation{University of Groningen, KVI, 9747 AA Groningen, The Netherlands}
\author{M. Sohani}
\altaffiliation[Present address: ]{Shahrood University of Technology, 36199-95161 Shahrood, Iran}
\affiliation{University of Groningen, KVI, 9747 AA Groningen, The Netherlands}
\author{D.J. van der Hoek}
\affiliation{University of Groningen, KVI, 9747 AA Groningen, The Netherlands}
\author{A. Rogachevskiy}
\affiliation{University of Groningen, KVI, 9747 AA Groningen, The Netherlands}
\author{E. Traykov}
\altaffiliation[Present address: ]{GANIL CEA/DSM-CNRS/IN2P3, Boulevard H. Becquerel, F-14076, Caen, France}
\affiliation{University of Groningen, KVI, 9747 AA Groningen, The Netherlands}
\author{O.O. Versolato}
\altaffiliation[Present address: ]{Max-Planck-Institut f\"ur Kernphysik, Saupfercheckweg 1, 69117 Heidelberg, Germany}
\affiliation{University of Groningen, KVI, 9747 AA Groningen, The Netherlands}
\author{L. Willmann}
\altaffiliation[Present address: ]{University of Groningen, FMNS, 9747 AG Groningen, The Netherlands}
\affiliation{University of Groningen, KVI, 9747 AA Groningen, The Netherlands}
\author{H.W. Wilschut}
\altaffiliation[Present address: ]{University of Groningen, FMNS, 9747 AG Groningen, The Netherlands}
\affiliation{University of Groningen, KVI, 9747 AA Groningen, The Netherlands}
\author{A.R. Young}
\affiliation{Triangle Universities Nuclear Laboratory (TUNL), Durham, NC, 27708  USA}
\affiliation{North Carolina State University, Raleigh, NC 27695  USA}

\date{\today}

\begin{abstract}
The $\frac{1}{2}^+\rightarrow\frac{1}{2}^+$ superallowed mixed mirror decay of $^{19}$Ne to $^{19}$F is excellently suited for high precision studies of the weak interaction. However, there is some disagreement on the value of the half-life. In a new measurement we have determined this quantity to be $T_{1/2} = 17.2832 \pm 0.0051_{(stat)} \pm 0.0066_{(sys)}$~s, which differs from the previous world average by 3 standard deviations. The impact of this measurement on limits for physics beyond the standard model such as the presence of tensor currents is discussed.

\end{abstract}


\maketitle

We can precisely characterize the structure of the weak interaction and probe for physics Beyond the Standard Model (BSM) in nuclear beta decay by searching for a deviation from unitarity of the Cabibbo-Kobayashi-Maskawa (CKM) matrix, and by constraining the presence of new, exotic interactions which would manifest as scalar or tensor couplings. Currently, the thirteen, purely vector, superallowed $0^+ \rightarrow 0^+$ Fermi decays provide the most precise determination of the CKM matrix element governing nuclear beta decay, $|V_{ud}| = 0.97425(22)$, and the most stringent constraint on scalar couplings from the constancy of the $\mathcal{F}t$ values \cite{HardyTowner0plus2009}. The superallowed, mixed Fermi/Gamow-Teller transitions between $T = \frac{1}{2}$ mirror nuclei can provide a needed cross check of the nucleus dependent theoretical corrections which contribute significantly to the overall uncertainty, and have been used to extract $|V_{ud}| = 0.9719(17)$ \cite{NavSevHalfPlus2009}, where the $^{19}$Ne system is the most precise component. The $^{19}$Ne system is capable of very sensitive tests for right-handed currents due to the very small beta asymmetry \cite{NavetaLR1991}, and we show that it can be used to obtain complementary limits on tensor couplings.

Mixed decays require a measurement of the half-life and an angular correlation to perform SM tests, in order to fix the Fermi/Gamow-Teller mixing ratio $\rho$. Until recently, the uncertainty in the $^{19}$Ne half-life determination was several times higher than all other contributions to the uncertainty in the $\mathcal{F}$t value, and was on par with the contribution of the beta asymmetry parameter to the uncertainty of $V_{ud}$. There is some disagreement between the published values of the lifetime, represented by a $\chi^2$/NDF of 41.2/7 for the global data set. This discrepancy has been addressed in our experiment through a novel approach which addressed with improved reliability all of the major systematic concerns encountered in previous experiments: the possibility of diffusion of the $^{19}$Ne, radioactive contaminants in the $^{19}$Ne counting sample, and rate dependent effects associated with deadtime and pile up corrections.

This measurement was performed at the TRI$\mu$P dual magnetic separator facility at KVI \cite{BergetalTrimp2006}. We employed the $^{19}$F(p,n)$^{19}$Ne reaction in inverse kinematics in a H$_2$ gas target cell \cite{TrayetalTarget2007} in order to filter the resulting $^{19}$Ne beam.  A 20~$\mu$m thick aluminum degrader was installed in the dispersion plane for additional isotope separation between magnetic stages. The incident beam energy of 10.5~MeV/A was selected to take advantage of the high cross section for $^{19}$Ne production and low expected yields of contaminants such as $^{15}$O and $^{17}$F \cite{Kitetalpn1990} while having sufficient energy to deposit at a depth that would enable us to quantitatively characterize the effect of diffusion. The particle beams produced in the target were studied using silicon detectors in the dispersion plane and at the exit of the separator, and analyzed using the \texttt{LISE++} fragment separator simulation program \cite{LISE2008} and \texttt{SRIM} (Stopping and Range of Ions in Matter) simulation software package \cite{ZiegetalSRIM2010}.

\begin{figure}
  \includegraphics[trim=4.8cm 3cm 4.4cm 1cm, clip=true,
 width=0.35\textwidth]{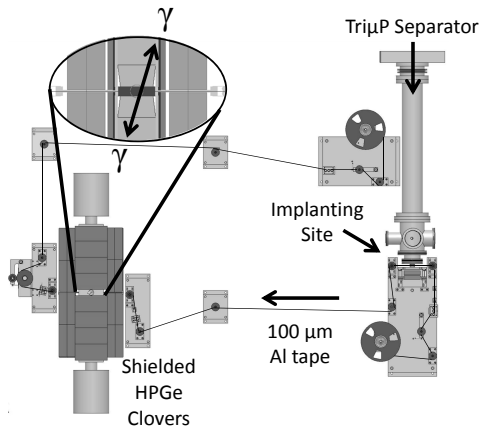}
  \caption{\label{fig:det} A reel to reel tape drive transported $^{19}$Ne samples from the TRI$\mu$P separator to the shielded detection system.
Two HPGe clovers detected the 511~keV annihiliation gammas in coincidence.}
\end{figure}

The particles were implanted up to 25~$\mu$m deep into a 100~$\mu$m thick aluminum tape. A custom tape drive system was used to transport the $^{19}$Ne sample from the isotope separator to the detectors (Fig.~\ref{fig:det}). The highest statistical sensitivity was achieved using an implanting duration of 50~s and a beam off counting time of 70~s. To study the effect of contaminants, an enhanced contamination mode was implemented using implanting and beam off counting durations of 240~s and 200~s, respectively. Counting began 20~s after the beam was turned off regardless of tape translation time.

The detection system was shielded by a graded assembly of aluminum, copper, steel, and lead. Two high purity germanium (HPGe) clover detectors \cite{Duchene199990} measured in coincidence the collinear 511~keV annihilation gamma rays from the decay positron, limiting the ambient background rate to $\sim$0.2~s$^{-1}$. The shaped clover signals were read by a CAEN V1724 waveform digitizer using the \texttt{MIDAS} data acquisition software package \cite{MIDAS}. Events were identified by comparing pulses to a precompiled average pulse shape. They were digitally filtered by a coincidence gate requiring at least one segment from each clover to have energy above 50~keV, and subjected to a deadtime of 5~$\mu$s. A pulser with constant repetition rate of 120~Hz was operated continuously for online monitoring of rate dependent effects.

The half-life was extracted from a blinded analysis described in detail in a PhD thesis \cite{Broussard2012}. A separate analysis was developed \textit{after unblinding} to cross check the systematic uncertainty. In the blind analysis, a factor of up to $\pm5\%$ was applied to the event timestamps before they were histogrammed into decay curves. The result was taken from the error weighted mean of half-lives extracted from the fits of the individual decay curves. The technique was studied using Monte Carlo simulation. An overall fitting bias of about 0.01$\%$ was found on the final extracted half-life due to the short fitting interval of 70~s.

A rate dependent factor $(1 + R(t)D)$ was included in the fitting function to account for a fixed deadtime $D$ on the counting rate $R(t)$. Time intervals containing pulser events or very high energy background events which saturated the detector were excluded from the fit. An additional bias was identified when the applied fixed deadtime and the deadtime due to the excluded intervals were simultaneously above a few percent. The coincidence gate for the two clovers was digitally varied from 0.3~$\mu$s to 1~$\mu$s and a correction for accidental coincidences was applied to the half-life by extrapolating to a gate width of 0~$\mu$s. Simulated data sets were used to estimate the uncertainty due to deadtime and accidental coincidences. Energy distortion due to pile up was mitigated by the clover segmentation. The rate of pile up over energy thresholds was calculated by sampling from the observed energy spectra, and the effect on the extracted half-life was negligible. The observed gain drift, after correcting for temperature variations, also had a negligible effect.  The uncertainty due to these effects was estimated to be $0.01\%$ from the variation in the extracted half-life due to energy thresholds.

\begin{figure}[]
  \includegraphics[width=0.45\textwidth]{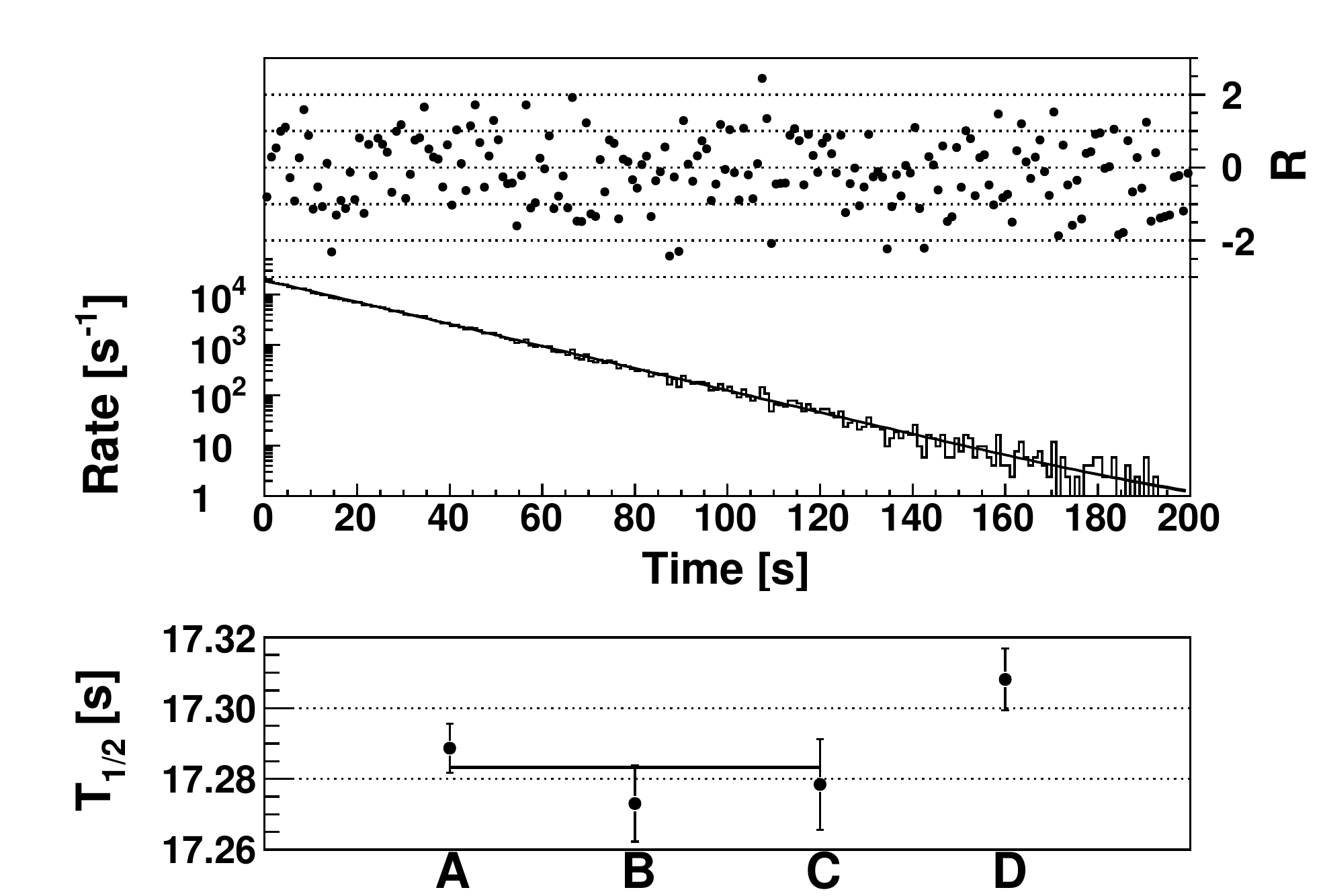}
  \caption{\label{fig:dc} (Top) Typical decay curve from enhanced contamination (200~s counting) data set with normalized residuals $R$. (Bottom) Half-life extracted from data sets with 50~s counting and no degrader (A), 9~$\mu$m degrader (B), and 18~$\mu$m degrader (C), and with 200~s counting, no degrader (D). The half-life is extracted from a fit of data sets A, B, and C to a model including the effect of diffusion.}
\end{figure}

$^{19}$Ne has been observed to diffuse very rapidly in Mylar, resulting in a shorter apparent half-life \cite{[][{ (unpublished)}]IacobDiffusion2011}. This effect was mitigated in the present measurement by implanting the sample into aluminum tape at an average depth of 25~$\mu$m, and was studied by reducing the depth of implantation so that the $^{19}$Ne diffused out of the tape more rapidly.  The implantation depth was reduced by placing either a 9 $\mu$m or 18  $\mu$m thick aluminum degrader immediately in front of the tape. The implanted depth profile as a function of degrader thickness was simulated using \texttt{SRIM}, and the diffusion of the particles in one dimension was numerically solved in a Monte Carlo simulation. The half-lives extracted from the 18~$\mu$m, 9~$\mu$m, and no degrader data sets were fit to this model to determine the best fit value for the half-life and diffusion coefficient (Fig.~\ref{fig:dc}). The extracted diffusion coefficient, $(1.00 \pm 0.52_{(stat)}) \times 10^{-2}~\mu$m$^2$s$^{-1}$, had a negligible effect on the half-life at all implantation depths. 

Limits on the contamination level in the sample were determined using the silicon detectors in the TRI$\mu$P separator and by using the HPGe clovers to search for characteristic gamma ray lines emitted by contaminant nuclei.  Several long lived positron emitting contaminants were not sufficiently constrained using these techniques. An additional data set was taken with 240~s implanting and 200~s counting, so that their concentration relative to $^{19}$Ne was significantly increased. The half-life extracted from this data set was $0.14 \pm 0.06\%$ higher (a 3~$\sigma$ shift) than the 50~s implanting, 70~s counting data sets. Monte Carlo simulations were used to determine the concentration of each \textit{single} contaminant required
to produce this observed increase. The expected shift due to $^{17}$F was taken as a conservative estimate for the uncertainty. Any combination of the longer lived contaminants would have resulted in a smaller effect on the extracted half-life.

\begin{table*}
  \caption {Statistical and systematic uncertainties in the blinded and post blind analysis.}
  \label{tab:final}
  \begin{ruledtabular}
    \begin{tabular}{|l|cc|cc|}\
      & \multicolumn{2}{c|}{Blind Analysis} & \multicolumn{2}{c|}{Post Blind Analysis (Full Data Set)} \\
      Source & Correction $(\%)$ & Uncertainty $(\%)$ & Correction $(\%)$ & Uncertainty $(\%)$ \\
      \hline
      Clock & - &  $1\times10^{-2}$ & - & $1\times10^{-3}$\\
      Fitting Bias & $-1.3\times10^{-2}$ & $8\times10^{-6}$ & - & $1\times10^{-3}$\\
      Deadtime\footnote{After unblinding, the uncertainty due to the deadtime correction was expanded to $0.02\%$ to include the shift in the half-life due to an error in implementation.} & $+3.0\times10^{-2}$ & $6\times10^{-3}$ & $+4.4\times10^{-2}$ & $2\times10^{-2}$ \\
      Accidental Coincidences & $-6.9\times10^{-2}$ & $2\times10^{-3}$ & $-6.0\times10^{-2}$ & $2\times10^{-3}$ \\
      Energy Determination & $+4.3\times10^{-2}$ & $1\times10^{-2}$ & $+5.6\times10^{-2}$ & $1\times10^{-2}$ \\
      Diffusion & $+4.8\times10^{-4}$ & $8\times10^{-5}$ & $+4.7\times10^{-4}$ & $7\times10^{-5}$ \\
      Ambient Backgrounds & $+5.2\times10^{-2}$ & $1\times10^{-3}$ & $+1.3\times10^{-1}$ & $4\times10^{-3}$\\
      Contamination & - & $3\times10^{-2}$ & $+2.8\times 10^{-2}$ & $3\times10^{-2}$\\
      Total Systematic &  $+4.3\times10^{-2}$ & $3.3\times10^{-2}$ & $+1.9\times10^{-1}$ & $3.7\times10^{-2}$ \\
      Total Statistical & & $2.9\times10^{-2}$ & & $2.5\times10^{-2}$ \\
      Result [s] & \multicolumn{2}{c|}{$17.2832\pm 0.0051_{(stat)} \pm 0.0058_{(sys)}$} & \multicolumn{2}{c|}{$17.2826 \pm 0.0044_{(stat)} \pm 0.0064_{(sys)}$} 
    \end{tabular}
  \end{ruledtabular}
\end{table*}

The final result for the half-life was extracted from the 50~s implantation data sets, including all three implantation depths. The corrections for statistical bias, deadtime, and accidental coincidences were applied to the half-life extracted from each measurement cycle. The unblinding factor was revealed to be $2.75\%$, so that the unblinded value for the half-life of $^{19}$Ne was $T_{1/2} = (17.2832 \pm 0.0051_{(stat)} \pm 0.0058_{(sys)} )$~s. The corrections and uncertainties for this result (``Blind Analysis") are summarized in Table \ref{tab:final}.

After unblinding, an alternate analysis was developed to confirm that the systematic shift in the $^{19}$Ne half-life was included by the uncertainty quoted in the blind analysis. Possible contaminants were included as parameters in a simultaneous fit of the short and long implantation data sets. A likelihood contour map covering the entire parameter space was constructed for each measurement cycle. The sum of the contour maps from all measurement cycles is equivalent to a simultaneous fit of the entire data set. The likelihood confidence intervals are constructed using $X^2 = - 2 \hspace{2pt} \text{ln} [ \frac{L(\theta)}{L_{max}} ] \leq \chi^2_{(1-\alpha)}$, where the quantity $X^2$ is comparable to $\Delta\chi^2$. No fitting bias was observed in the best fit parameters. This approach is fundamentally equivalent to the blind analysis method, and both methods extracted identical results in simulation and in the data. 

The approach for correcting accidental coincidences was altered to accommodate the simultaneous fit. The accidentals rate was calculated from the digitally varied coincidence window and included in the fitting function. This method was directly compared to the blind approach in simulation and in the data, and the results were in agreement at better than the $10^{-5}$ level. 

The contaminant isotopes which could not be otherwise excluded, $^{17}$F, $^{18}$F, and $^{15}$O, were included as fit parameters. $^{13}$N and $^{11}$C were not included because the total concentrations were low enough, and their half-lives were such that that they would not be distinguished from the $^{15}$O and $^{18}$F populations. The fitting procedure was demonstrated in simulations to correctly extract the half-life for the low concentrations of contaminants expected in the experiment. The $X^2 = 1$ contour overestimated the actual uncertainty (68$\%$ confidence interval) by about 20$\%$, but was adopted as it is more conservative. The best fit values for the relative concentrations of $^{15}$O, $^{17}$F, and $^{18}$F from the simultaneous fit of the full data set were found to be 0.003$\%$, 0.000$\%$, and 0.030$\%$ of $^{19}$Ne, which agreed with the limits from the silicon detector at the separator exit, and the limits from the blinded analysis. 

We note that post unblinding, we encountered an error in the implementation of the correction due to deadtime, which incorrectly increased the extracted half-life by 0.02$\%$.
We did not correct the implementation, but instead increased our uncertainty to accommodate that shift. The systematic uncertainties for the blind and post blind analyses are tabulated in Table~\ref{tab:final}. The small variations are due to the different sensitivities to systematic effects of the expanded fit intervals in the additional data set.

\begin{figure}
  \includegraphics[trim=0cm 0.5cm 0cm 0.5cm, clip=true, width=0.35\textwidth]{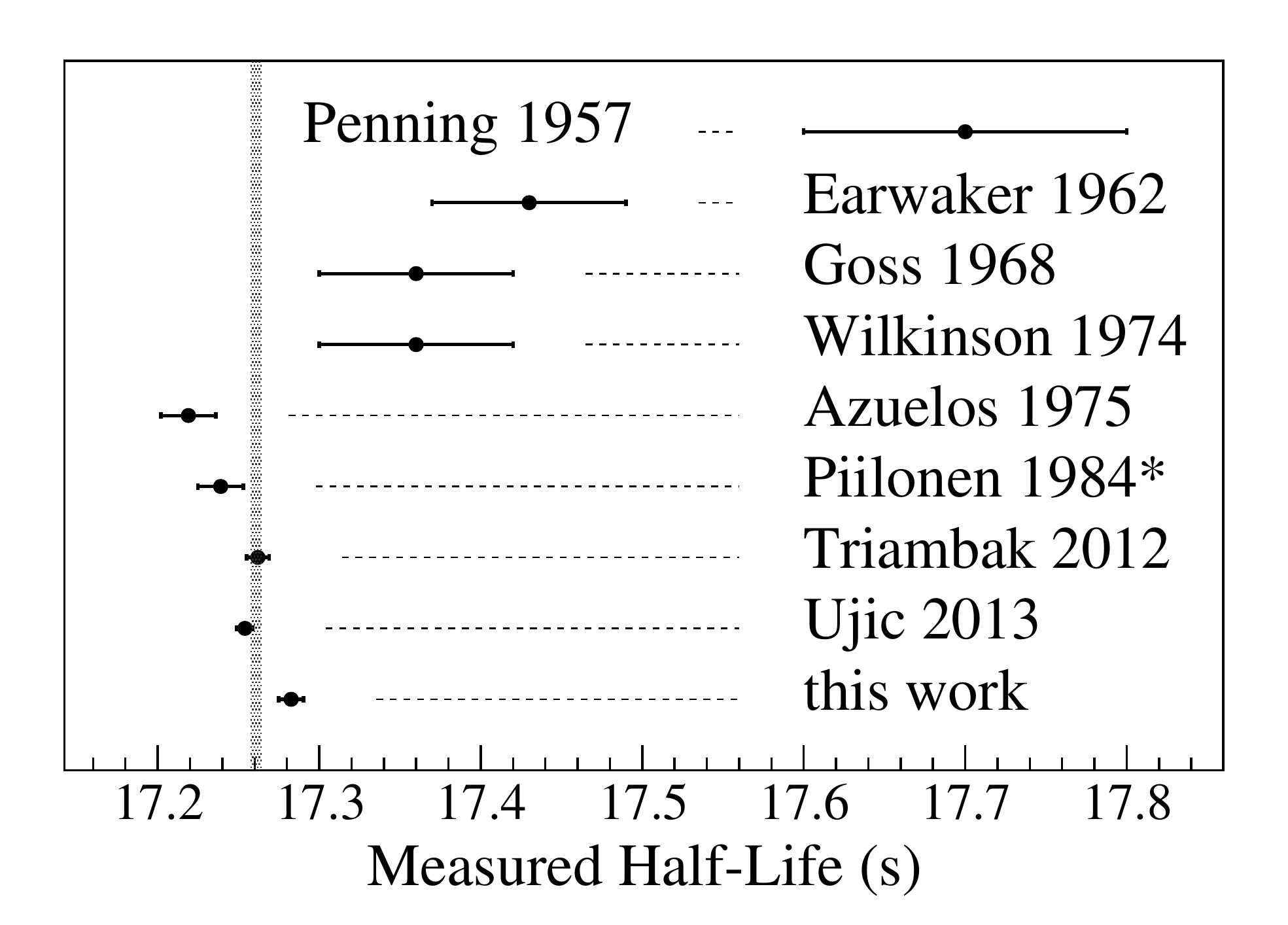}
  \caption{\label{fig:newave} Previous measurements of the half-life, performed by Penning \cite{PennTau1957}, Earwaker \cite{ EarwTau1962}, Goss \cite{GossTau1968}, Wilkinson \cite{WilkAlbTau1974}, Azuelos \cite{AzKitchTau1975}, Piilonen (not published) \cite{PiilTau1984}, Triambak \cite{ TriambakTau2012}, and Uji\'{c} \cite{ UjicTau2013}.
The error weighted average is $T_{1/2} = (17.2604\pm0.0034)$~s (vertical line).}
\end{figure}

The ``post blind" analysis allowed for contaminant half-lives to shift the central value of the half-life relative to the blinded result, included the additional statistics of the enhanced contamination data set, and expanded the systematic uncertainty.  We obtained $T_{1/2} = (17.2826 \pm 0.0044_{(stat)} \pm 0.0064_{(sys)} )s$, which is consistent with the result from our blinded analysis.

We now include the result for the half-life of $^{19}$Ne from the blinded analysis, with expanded systematic uncertainty, of $T_{1/2} = 17.2832\pm0.0051_{(stat)}\pm0.0066_{(sys)}s$, with the previous measurements \cite{PennTau1957, EarwTau1962, GossTau1968, WilkAlbTau1974, AzKitchTau1975, PiilTau1984, TriambakTau2012, UjicTau2013} and obtain the error weighted average (Fig.~\ref{fig:newave}), $T_{1/2} = (17.2604\pm0.0034) s$.  The $\chi^2$/NDF of the fit, 50.3/8, indicates the presence of unaccounted systematic effects. This situation is highlighted in a comparison between the approaches of the most precise, published experiments, where our result differs from Triambak {\it et al.}~\cite{TriambakTau2012}  by 2 $\sigma$, and from Ujic {\it et al.}~\cite{ UjicTau2013} by 3 $\sigma$. Our measurement is not sensitive to the afterpulsing and gain shifts associated with the scintillator and PMT assemblies used in Triambak {\it et al.} and Ujic {\it et al.}, but our measurement does incorporate larger corrections due to rate-dependent effects, especially dead time, and potential contaminant effects.  Our methodology also differed in a number of respects from both of the other recent experiments, including the direct characterization of the possible influence of diffusion and contaminants, which provides a solid point of comparison to these other measurements. $^{19}$Ne is of particular interest for its potential to provide useful constraints on BSM physics, motivating further attention to developing a robust data set for the half-life and decay parameters of this system.

We can calculate $\mathcal{F}t= 1719.8(13)$ s as defined by Ref.~ \cite{SevTownFtVals2008} from the error weighted average of all available experimental results, which is combined with the  value $\rho = 1.5995(45)$ obtained from the beta asymmetry  $A_0 = -0.0391(14)$ \cite{CaletalAbeta1975}, to extract $V_{ud}= 0.9712(22)$. This value is more precise than that obtained from any mirror decay other than the neutron,
for which the PDG 2013 values for the lifetime and asymmetry lead to a value of $V_{ud} = 0.9774(17)$ \cite{PDG2012}.

The $^{19}$Ne system can also be used to extract limits on tensor couplings, without recourse to typical multiparameter fit procedures, if one utilizes the vector coupling strength derived from superallowed $0^+ \rightarrow 0^+$ decays \cite{plasterva2013,GarciaTensor2013}. We follow a procedure similar to one developed to extract tensor limits in neutron decay, described in detail in Ref.~ \cite{PattieTensor2013}, in which we include the contribution of a Fierz interference term to the decay rate and the measured beta asymmetry. However, we also include the contribution to the expected SM \textit{energy dependence} of the asymmetry, in a reanalysis of the data set of Ref.~ \cite{CaletalAbeta1975}, in determining $\rho$. We extract the 2 $\sigma$ (95$\%$ C.L.) limits on the Fierz term, $ -0.050 < b < 0.007$, and the corresponding limits on BSM tensor currents, $-0.006 < \frac{C_T}{C_A} < 0.034$ (95$\%$ C.L.). This result is about a factor of 10 less precise than the recent limits determined from a data set which includes essentially all relevant nuclear beta decays and neutron decay \cite{GarciaTensor2013}. This analysis will be further developed in a future publication.

The $^{19}$Ne system has the potential for significant improvements and an impact comparable to that of the neutron. Improved constraints on BSM physics depend critically on the angular correlation data available for $^{19}$Ne. More precise studies of the energy dependence of the beta asymmetry and electron neutrino correlation and a careful analysis of the current experimental status of the constraints on scalar and tensor currents is warranted. Improvement in the determination of $\rho$ to similar precision as the $\mathcal{F}t$ value would both reduce the uncertainty in $V_{ud}$ and the limits on $b$ by a factor of nearly 2, or more if  uncertainties in the ratio of $\frac{f_A}{f_V}$ are addressed. This would make the value of $V_{ud}$ extracted from $^{19}$Ne decays more precise than the value determined from neutron decays, defining the cutting edge on a meaningful cross check for the value of $V_{ud}$  from the superallowed $0^+ \rightarrow 0^+$ Fermi decays.

\begin{acknowledgments}
We wish to thank all of the technical staff who made this measurement possible, especially M. Busch, B. Carlin, J. Faircloth, L. Huisman, and P. Mulkey. We are grateful to I.S. Towner for helpful discussion and calculations of $\left<\frac{1}{W}\right>$. This work was supported by the U.S. National Science Foundation (grant NSF-1005233) and the U.S. Department of Energy (grant DE-FG02-97ER41042). This work is part of the research programmes 48 and 114 of the Foundation for Fundamental Research on Matter (FOM), which is part of the Netherlands Organisation for Scientific Research (NWO).
\end{acknowledgments}

\bibliography{mybib2}

\end{document}